\documentclass[a4paper,11pt]{article}
\usepackage{latexsym}
\usepackage{feynmf}		
\def\ba{\begin{eqnarray}}
\def\ea{\end{eqnarray}}

\def\lb{\label}
\def\be{\begin{equation}}
\def\ee{\end{equation}}

\begin{document}
\baselineskip0.25in

\title{Cosmological singularity theorems for $f(R)$  gravity theories}
\author{Ivo Alani \thanks{Instituto de F\'isica de Buenos Aires (IFIBA)} and
Osvaldo P. Santill\'an \thanks{Departamento de Matem\'aticas Luis Santal\'o, Buenos Aires, Argentina
firenzecita@hotmail.com and osantil@dm.uba.ar.} }
\date {}
\maketitle

\begin{abstract}
In the present work  some generalizations of the Hawking singularity theorems in the context of $f(R)$ theories are presented. The assumptions  of these generalized theorems is that the matter fields satisfy the conditions 
$\bigg(T_{ij}-\frac{g_{ij}}{2} T\bigg)k^i k^j\geq 0$ for any generic unit time like field $k^i$, that the scalaron takes bounded positive values during its evolution, and that the resulting space time is globally hyperbolic.
Then, if there exist a Cauchy hyper surface $\Sigma$ for which the expansion parameter $\theta$ of the geodesic congruence emanating orthogonally from $\Sigma$ satisfies some specific conditions, it may be shown 
that the resulting space time is geodesically incomplete. Some mathematical results of reference \cite{fewster} are very important for proving this. The generalized theorems presented here apply directly some specific models such as the Hu-Sawicki or Starobinsky ones \cite{especif3}, \cite{capoziello4}. For other scenarios, some extra assumptions should be 
implemented for the geodesic incompleteness to take place. However, the negation of the hypothesis of these results does not necessarily  imply that a singularity is absent, but that other mathematical results should be considered in order to prove that.
\end{abstract}

\section{Introduction}
One of the most interesting problems in cosmology is the accelerated expansion of the universe \cite{acelerado}. Some models for explaining
this acceleration are quintessence mechanisms \cite{quintessence}-\cite{quintessence2}. These models include some  components which induce a repulsive term imitating the effect of a cosmological
constant \cite{cosmological constant}.  Another problem of importance in cosmology arises when studying the behavior of the galactic rotation curves and
the mass discrepancy in clusters of galaxies \cite{rotation}.  The observations show
that the mass increases linearly with the radius near the galaxy center. 
The estimated baryonic mass density seems not enough for explaining this, and the mass discrepancy is usually explained by postulating the existence
of a dark matter.  This component is assumed to be cold and pressureless.
Several dark matter scenarios include weakly interacting massive particles
(WIMP) \cite{wimp}-\cite{wimp3}. These components may be light or superheavy, but 
should interact weak enough with ordinary matter in order not to be detected by the current accelerator technology.

A further  possibility to explain these experiments is that GR breaks down at a typical galaxy scale.  In fact, there exist observations
that suggest deviations from the standard General Relativity \cite{will}. Modified gravity theories and, in particular, $f(R)$ theories, are a plausible alternative. From the theoretical point of view, deviations of GR
were considered long time ago. For instance, corrections of the form $f(R)\sim R+\alpha R^2$ have been shown to cast  an early inflationary period \cite{especif2}, while $f(R)\sim R+\alpha R^{-1}$ can give rise to a present
accelerated period \cite{especif3}. However, for functions $f(R)$ in powers of $R$, it has been shown that the resulting scale factor has a time dependence of the form $a(t)\sim t^{1/2}$ instead of $a(t)\sim t^{2/3}$ for small and large curvature \cite{cosmos1}-\cite{cosmos3}. Nevertheless, these theories can cast a matter dominated period followed by an accelerated expansion, and although several models have been discarded, viable models seem to exist \cite{capoziello3}-\cite{cut8}. Several phenomenological aspects of these models were discussed recently in \cite{tegmark}, which special emphasize in observations.

A serious problem related to curvature singularities in $f(R)$ scenarios was pointed out in  \cite{frolov1}-\cite{frolov2} and \cite{maeda}.  As is well known, $f(R)$ theories reduce, by means of a suitable conformal transformation, into GR plus an scalar field $\varphi$, known as the scalaron. This scalar is under the action of a potential term $U(\varphi)$, and couples non trivially to the matter fields. The problem is that, for $f(R)$ theories which reduce to GR in the large $R$ limit, the potential term corresponds to an unprotected curvature singularity. The presence of such singularities
has disastrous observational consequences, in particular it may not allow the formation of relativistic stars such as neutron ones. However, there exist some models which seems to avoid these problems and may be realistic as well, such as the ones presented in \cite{maeda2}.

The present work is related to singularities in $f(R)$ gravity theories in the cosmological context. The formulation of these theories this work is concerned with is the metric one, and the resulting equations for the space
time metric are of fourth order. In particular, there is an extra scalar degree of freedom in this formulation, known as the scalaron. The definition of singularity that it will be adopted here is the Geroch one namely, past or future geodesic incompleteness.  The main goal is to present some singularity theorems which generalize the Hawking cosmological theorems, in the context of $f(R)$ theories. This generalization present some technical complications, since the standard matter field condition $\bigg(T_{ij}-\frac{g_{ij}}{2} T\bigg)k^i k^j\geq 0$ for any generic unit time like field do not imply the condition $R_{ij}k^i k^j\geq 0$ for $f(R)$ theories, and the last condition is of essential importance in the proof of the Hawking theorems. In addition, the scalaron degree of freedom spoils the first condition. The aim of the present work is to sort out these problems and formulate some concrete generalization of the Hawking results. Some mathematical results proven in \cite{fewster} are crucial for this achievement.

The present work may have an overlap with some appearing in the literature. For instance, in \cite{dadum3} the finite future  
singularities emerging in alternative gravity dark energy models were studied both in the Jordan and Einstein frames. These reference shows that a singularity can be present even for a flat space time. Some viable
modified gravity models were analyzed in detail and a cure to a singularity was found in terms of higher order curvature corrections.
Also, in reference \cite{dadum1}  the future evolution of the dark energy in modified gravity theories including
$f(R)$ gravity, string-inspired scalar-Gauss-Bonnet and modified Gauss-Bonnet ones were studied. Several examples
of the modified gravity which produces accelerating cosmologies ending at the finite-time future
singularity were found and some scenarios to
resolve the finite-time future singularity were presented. A natural scenario is
is related with additional modification of the gravitational action in the early universe.  Furthermore, it is shown that the non-minimal gravitational coupling can remove the
finite-time future singularities or make the singularity stronger (or weaker) in modified gravity. Further features are reviewed in \cite{dadum2}.

The present work is organized as follows. In section 2 some generalities about $f(R)$ theories are reviewed, in particular the role of the scalaron. In section 3 the standard Hawking results are reviewed, and some important fact about strong energy conditions is reminded. Section 4 is devoted to some properties of the Raychaudhuri equation, which are further applied to obtain our singularity results. These results are applied to some known $f(R)$
theories and are all related to compact Cauchy hyper surfaces. Section 6 contains some results that allows to relax the mentioned compactness assumption. Section 7 contains a discussion about the initial data formulation for these models. Section 8 contains a discussion of the obtained results and open problems to be considered further.

\section{Generalities about $f(R)$ theories}
The signature convention to be adopted in the following is $(-,+,+,+)$. As mentioned in the introduction, the $f(R)$ models are generalizations of the GR. We collect here some basic features, further information can be found
in the extensive references \cite{capoziello3}-\cite{cut8}.
These theories start with the following lagrangian 
\be\lb{fr}
S = \frac{1}{16\pi G_N} \int f(R) \sqrt{-g}d^4x+S_m,
\ee
where the matter action $S_m$ for a given component takes the same form as in GR. 
This class of theories reduce to the Einstein GR when $f(x)=x$. 

One of the main differences between $f(R)$ models and the standard GR 
is that, for the former, the Palatini or metric formalism give the same theory. Recall that the Palatini formalism involves an independent variation of the metric and the connection, 
while for the metric formalism the metric is the only object to be varied. For GR, both formalisms give the same system of equations, while for $f(R)$ theories
both systems are inequivalent. The following exposition will be centered in the metric formalism. The
corresponding generalized Einstein equations are of fourth order, and their explicit form is given by
\be\lb{eqm}
f'(R)R_{ij}-\frac{1}{2}f(R)g_{ij} - [\nabla_i\nabla_j - g_{ij}g^{ab}\nabla_a\nabla_b]f'(R) = 8\pi G_N T_{ij}.
\ee
The energy momentum tensor $T_{ij}$ introduced in the left hand side is the standard one 
\be\lb{em}
T_{ij}=-2\frac{\delta {\cal L}_m}{\delta g^{ij}}+g_{ij}{\cal L}_m,
\ee
with ${\cal L}_m$ the matter lagrangian defined through
$$
S_m=\int  {\cal L}_m \sqrt{-g}d^4x.
$$
The equations (\ref{eqm}) become of second order only in the specific situation $f(x)=x$, which corresponds to the GR limit. 
In other situations, the contraction of these equations with the inverse metric tensor $g^{\mu\nu}$ shows that
$$
f'(R)R - 2 f(R) + 3 g^{ab}\nabla_a\nabla_b f'(R) = k T,
$$
which is a differential equation for $f(R)$ unless $f(x)=x$. This is a non algebraic equation, and suggest that
that $f(R)$ can be interpreted as an additional degree of freedom.  In order to clarify this, it is convenient to introduce the following formulation of the theory
\be\lb{verv}
S=\frac{1}{16\pi G_N} \int (R \phi - V(\phi))\sqrt{-g}d^4x+S_m,
\ee
where $\phi$ is an auxiliary field, since it has no kinetic term. The lagrangian ${\cal L}(\phi, R)=R \phi - V(\phi)$ should be identified as $f(R)$. Therefore $f(R)$ is the Legendre transform of $V(\phi)$.
From the equations of motion for $\phi$ and some standard properties of Legendre transforms it follows that
\be\lb{dufo}
R=V'(\phi),\qquad \phi=f'(R),
\ee
The equations of motion resulting from (\ref{verv}) may be simplified by making the following conformal transformation (to the Einstein frame)
\be\lb{ct}
\phi \longrightarrow \varphi = \sqrt{\frac{3}{2k}} \log \phi,\qquad 
g_{ab} \rightarrow \widetilde{g}_{ab} = \phi g_{ab}, 
\qquad
\sqrt{-g} = \phi^{-2} \sqrt{-\widetilde{g}}.
\ee
The inverse transformations are
 \be\lb{ct2}
\varphi \longrightarrow \phi=e^{\sqrt{\frac{2k}{3}} \varphi} ,\qquad 
\widetilde{g}_{ab} \rightarrow g_{ab} = e^{-\sqrt{\frac{2k}{3}} \varphi} \widetilde{g }_{ab}, 
\qquad
\sqrt{-\widetilde{g}}= e^{\sqrt{\frac{8k}{3}} \varphi} \sqrt{-g}.
\ee
By taking into account the well known transformation of the curvature scalar $R$ under conformal transformation
\be\lb{cam}
\phi\widetilde{R}=R-\frac{3}{2}g^{\mu\nu}\partial_\mu\log\phi \partial_\nu\log\phi-3\Box \log\phi,
\ee
and by introducing the last expression into (\ref{verv}),   it is obtained after neglecting a total derivative term the following action
\be\lb{eins}
S'_g[\widetilde{g}_{ab}, \varphi] =  \int \left[ \frac{\widetilde{R}}{2k} - \frac{1}{2} \widetilde{g}^{ij} \partial_i \varphi \partial_j \varphi - U(\varphi) \right] \sqrt{-\widetilde{g}} d^4x+S_m.
\ee
Here
\be\lb{pot}
U(\varphi) = \frac{V(\phi)}{2\kappa\phi^2} =\frac{ R f'(R)-f(R)}{2\kappa f'(R)^2},
\ee
and the matter lagrangian in $S_m$ should be related to the new frame. An elementary calculation shows that the matter lagrangian ${\cal \widetilde{L}}_m$ in the new frame is related to 
${\cal L}_m$ by
\be\lb{lm}
{\cal \widetilde{L}}_m(X, \widetilde{g}, \varphi)=\frac{1}{\phi^{2}} {\cal L}_m(X,  \widetilde{g}\phi).
\ee

The field $\varphi$ is known as the scalaron. Thus the $f(R)$ system has been reduced to GR coupled to the scalaron $\varphi$ plus generic mater fields $X$ interacting with the scalaron as well. 

\section{Strong energy conditions and cosmological singularity theorems}

The aim of the present work is to study possible generalizations of the GR singularity theorems for the $f(R)$ models discussed above.
As is well known in GR, an special role in studying singularity theorems is played by matter whose energy momentum tensor satisfy  the condition
\be\lb{sec}
\bigg(T_{ij}-\frac{g_{ij}}{2} T\bigg)k^i k^j\geq 0,
\ee
where $k^i$ is any  generic  time like field of unit length. For GR this implies the following inequality for the curvature tensor
\be\lb{sec2}
R_{ij}k^i k^j\geq 0,
\ee
which is known as the strong energy condition. On the other hand, given such vector field $k^i$, it can be interpreted as a congruence of non intersecting world lines, not necessarily geodesics.

The following discussion is restricted to a generic globally hyperbolic space time ($M$, $g$), with a particular choice of a Cauchy hyper surface $\Sigma$ where the initial data is formulated. Consider the geodesic congruence orthogonal to $\Sigma$. As is well known, the expansion scalar  $\theta_\gamma$ of this congruence satisfies the Raychaudhuri equation
\be\lb{raycha2}
\frac{d\theta_\gamma}{d\tau}=-\frac{\theta_\gamma^2}{n-1}-\textrm{Ricc}(\gamma', \gamma')-2\sigma^2,\qquad \theta_\gamma(0)=\theta_{\gamma0}.
\ee
Here $n$ is the space time dimension, $\textrm{Ricc}(\gamma', \gamma')=R_{ab} \gamma'^a\gamma'^b$ and $\gamma:[0, \infty)\to M$ is any of the future complete unit speed geodesic of the congruence, which emanates orthogonally from $\Sigma$. The vectors $\gamma'^a$\ are assumed to be of unit length and the geodesics are parameterized by the proper time $\tau$.

The expansion scalar $\theta_\gamma$ described by (\ref{raycha2}) has a fundamental geometrical interpretation. Given two points $p$ and $q$  in the space time ($M$, $g$), they are named conjugate points to each other when there is a non zero Jacobi field which vanishes both at $p$ and $q$. A necessary and sufficient condition for $q$ to be conjugate to $p$ is that, for the geodesic congruence emanating from $p$,  the expansion $\theta_\gamma\to-\infty$ at $q$. This notion can be extended to space like surfaces, such as $\Sigma$. A point $q$ on a geodesic $\gamma$ of the geodesic congruence orthogonal to  $\Sigma$ is said to be conjugated to it, if there is a Jacobi field which does not vanish at $\Sigma$ but vanish at $q$. Intuitively, the point $q$ is conjugated to $\Sigma$ when there are two infinitesimally close geodesics emanating from $\Sigma$ which converge to $q$. The divergence of $\theta_\gamma$ corresponding to the geodesic congruence will also diverge at $q$ in this situation  \cite{wald}, \cite{hawking}.

The conjugate points discussed above characterize the presence of a singularity.  Recall that in the signature $(-, +, +, +)$ we are working with, and for globally hyperbolic space time, there exist a  time like curve joining two points $p$ and $q$ which maximizes the proper time.  This curve is a geodesic without conjugate points between $p$ and $q$, and is maximizing along all the continuous time like curves connecting $p$ and $q$, not only the differentiable ones \cite{wald}, \cite{hawking}. In these terms it is possible to prove the following generic singularity theorem.\\

\emph{Proposition 1:} Consider a globally hyperbolic space time ($M$, $g$) with a given Cauchy surface $\Sigma$. If any solution  $\theta_\gamma$ of the Raychaudhuri equation (\ref{raycha2}) explodes
before  finite proper time $\tau_1>0$ (a condition which is to be developed further), then the space is future geodesically incomplete.
\\

Recall that a Cauchy hyper surface is achronal, and it is assumed that $\Sigma$ corresponds to the proper time $\tau=0$.
\\

\emph{Proof:} Assume that there exists a future directed time like curve $\lambda$ with length  $\tau>\tau_1$ in $M$. Let $q$ a point lying on $\lambda$ beyond the length $\tau_1$. For any globally hyperbolic space time there exist a curve $\gamma$ which maximize the proper time between $\Sigma$ and $q$ \cite{wald}, \cite{hawking}.  Let $p$ the intersection between this curve $\gamma$ and the Cauchy hyper surface $\Sigma$.  By its definition, the curve $\gamma$ maximize the proper time between $p$ and $q$. Thus, it should be a geodesic without conjugate points between $p$ and $q$. However, by the theorem assumptions, the expansion parameter $\theta_\gamma$ explodes at a finite proper time $\tau<\tau_1$ and thus there is a conjugate point between $p$ and $q$. This contradiction shows that the curve $\lambda$ does not exist and the space is future geodesically incomplete. (Q. E. D)
\\

The generalization of the previous theorem to past directed curves presents no difficulties at all. It is important to remark that the Raychaudhuri equation (\ref{raycha}) is purely geometric, and its deduction does not take into account wether the geometry is described by the Einstein equations or other type of gravity model. In addition, when the  condition (\ref{sec2}) is satisfied,  the second term of the right hand side is positive. Thus 
\be\lb{sim}
\frac{d\theta_\gamma}{d\tau}+\frac{\theta_\gamma^2}{n-1}\leq0,\qquad \theta_\gamma(0)=\theta_{\gamma0},
\ee
which implies that
$$
\frac{1}{\theta_\gamma}\geq\frac{1}{\theta_{\gamma0}}+\frac{1}{n-1}\tau.
$$
Clearly, this implies that when $\theta_{\gamma 0}>0$ in all the Cauchy surface, then $\theta_\gamma\to-\infty$ at a finite proper time $\tau\leq(n-1)/C$. Thus, the following theorem follows \cite{hawking}.
\\

\emph{Hawking singularity theorem (future version):} Consider an hyperbolic space time ($M$, $g$) with a Cauchy hyper surface $\Sigma$ for which
 $\theta_0(p)\geq C>0$ for every point $p$ at that surface. If the Ricci tensor satisfy the condition (\ref{sec2}) then the space time
is future geodesically incomplete. More precisely, all the future directed curves have length with absolute value no larger than $\tau_0=(n-1)/C$. 
\\

The past version of this theorem is straightforward.
\\

\emph{Hawking singularity theorem (past version):} Consider an hyperbolic space time ($M$, $g$) with a Cauchy hyper surface $\Sigma$ for which
$\theta_0(p)\leq C<0$ for every point $p$ at that surface. If the Ricci tensor of the geometry satisfy the condition (\ref{sec2}) then space time
is past geodesically incomplete. More precisely, all the past directed curves have length with absolute value no larger than $\tau_0=(n-1)/C$. 
\\

It is important to remark that this theorem is valid when the conditions (\ref{sec2}) are satisfied, independently on the gravity model describing the Ricci tensor $R_{ij}$.
An example for which it applies is GR with matter satisfying the strong energy condition (\ref{sec}), since this automatically implies (\ref{sec2}).
This simple argument do not apply for $f(R)$ theories even when the matter content satisfies (\ref{sec}). This is due to the fact that the equation (\ref{eqm}) does not imply  that (\ref{sec2}) is satisfied.  Thus, it is not trivial to understand for which situations $\theta_\gamma\to-\infty$ at finite time for $f(R)$ models, which implies geodesic incompleteness by the proposition 1.

Although  these drawbacks, there is an interesting observation to be made. By taking into account (\ref{lm}) it follows that the energy momentum for matter (\ref{em}) in the Einstein frame is 
\be\lb{em2}
\widetilde{T}_{ij}=\frac{1}{\phi}T_{ij},\qquad \widetilde{T}\widetilde{g}_{ij}=(\widetilde{T}_{ab}\widetilde{g}^{ab})\widetilde{g}_{ij}=\frac{1}{\phi}T g_{ij},\qquad \widetilde{g}_{ij}k^ik^j=-\phi
\ee
Thus, when the matter energy momentum tensor satisfies the strong energy conditions (\ref{sec}) in the Jordan frame, it also satisfy them in the Einstein frame if $\phi> 0$, namely
\be\lb{secu}
\bigg(\widetilde{T}_{ij}-\frac{\widetilde{g}_{ij}}{2} \widetilde{T}\bigg)\widetilde{k}^i \widetilde{k}^j\geq 0,
\ee
with $\widetilde{k}^i=\phi^{-1/2}k^i$ a generic  time like field of unit length defined on ($\widetilde{M}$, $\widetilde{g}$).  This  suggest the possibility of studying the time geodesically completeness of the conformal metric $\widetilde{g}$ defined in (\ref{ct}) and the properties of the scalaron $\varphi$ first, and then to analyze these matters for the Jordan metric $g$ later on.

The conditions (\ref{secu}) are encouraging. However,  it should
be kept in mind that the scalaron $\varphi$, as a generic scalar field, does not satisfy the strong energy condition (\ref{sec}). Thus, further work should be made in to avoid this problem.  We turn the attention to some specific theorems which, under certain circumstances, insure that the Einstein metric $\widetilde{g}$ geodesically incomplete when the matter field satisfy (\ref{sec}). After this point is clarified, we will deal with the singularities of the Jordan metric $g$, which is our main goal.

\section{Generalized singularity theorems for compact Cauchy hypersurfaces}
\subsection{General statements}
The previous section shows that a fundamental tool for studying singularity theorems is  the Raychaudhuri equation
\be\lb{raycha}
\frac{d\theta_\gamma}{d\tau}=-\frac{\theta_\gamma^2}{n-1}-\textrm{Ricc}(\gamma', \gamma')-2\sigma^2,\qquad \theta_\gamma(0)=\theta_{\gamma0}.
\ee
Some properties of its solutions may be deduced by making the change of variables $y(\tau)=-(\theta_\gamma+c_\gamma)e^{-2c_\gamma\tau/(n-1)}$ for which the last equation takes the form
\be\lb{teor}
\frac{dy}{d\tau}=\frac{y^2}{q(\tau)}+p(\tau),\qquad y(0)=y_0,
\ee
with \be\lb{che} q(\tau)=(n-1)e^{-2c_\gamma\tau/(n-1)},\qquad p(\tau)=e^{-2c_\gamma\tau/(n-1)}\bigg(\textrm{Ricc}(\gamma', \gamma')+2\sigma^2+\frac{c_\gamma^2}{n-1}\bigg).\ee
Equations of the type (\ref{teor}) were considered in the literature \cite{fewster} for a wide variety of functions $p(\tau)$ and $q(\tau)$, and some general knowledge
about them is the following. 
\\

\emph{Proposition 2:} Consider an equation of the form (\ref{teor}), and suppose that  the functions $p(\tau)$ and $q(\tau)$
satisfy
\be\lb{cudo}
\int_0^\infty \frac{d\tau}{q(\tau)}=\infty,\qquad \lim_{T\to\infty} \textrm{inf}\int^T_0 p(\tau)d\tau>-y_0,
\ee
with $q(\tau)>0$ in $[0,\infty)$.  Then, any  of the solutions $y(\tau)$ of (\ref{teor})  with the initial condition $y(0)=y_0$ do not extend to the interval $[0,\infty)$. In other words, the solution $y(\tau)$ explodes for a  time $\tau_0<\infty$.

\emph{Proof:} Assume on the contrary that $y(\tau)$ extend to the whole interval $[0,\infty)$. This will imply a contradiction which will show
that this affirmation is false. To see that, note that second assumption (\ref{cudo}) implies the existence of a time $\tau_1$ for which
$$
\int^t_0 p(\tau')d\tau'>-y_0, \qquad \textrm{for}\qquad \tau>\tau_1.
$$
By integrating the equation (\ref{teor}) and taking into account the last inequality, it follows that
\be\lb{teor2}
y(\tau)=\int_0^\tau \frac{y^2}{q} d\tau' +\int_0^\tau p d\tau'+y_0>\int_0^\tau \frac{y^2}{q} d\tau'.
\ee
Now, let us introduce the quantity given by
$$
R(\tau)=\int_0^\tau \frac{y^2(\tau')}{q(\tau')} d\tau'.
$$
As $q(\tau)$ is positive in the half positive line and $\tau>0$, it can directly be seen that this quantity is always positive.
This definition and the inequality (\ref{teor2}) shows that 
\be\lb{egoc2}
\frac{R^2}{q}<\dot{R}=\frac{y^2}{q},
\ee
for $\tau>\tau_1$. From here it is concluded, for every $\tau_2>\tau_1$, that
$$
 \int_{\tau_2}^{\tau}\frac{d\tau'}{q}<\int_{\tau_2}^{\tau}\frac{\dot{R}}{R^2}d\tau'=\frac{1}{R(\tau_2)}-\frac{1}{R(\tau)}<\frac{1}{R(\tau_2)}.
$$
However, the first condition (\ref{cudo}) it follows that the left hand is not bounded when $\tau\to \infty$. Thus, the last inequality makes sense only for times $\tau<\tau_0$, with $\tau_0$ a fixed time. This shows that 
$y(\tau)$ can not extend to the whole interval $ [0,\infty)$, but only to an interval inside $[0, \tau_0]$. (Q. E. D)
\\

Note that the last proposition predicts the existence of a finite time $\tau_0>0$ for which the solution $y(\tau)$  but it does not give an estimate about it. 
Now, the first condition (\ref{cudo}) applied to the equation (\ref{che}) simply give that $c_\gamma>0$. The second (\ref{cudo}) is translated after an elemental integration
into the following proposition \cite{fewster}.\\

\emph{Generalized future singularity theorem:} Consider a globally hyperbolic space time $M$ with dimension $n>2$ and let $\Sigma$ a compact Cauchy 
surface for it. Suppose that for each future directed time geodesic $\gamma: [0, \infty)\to M$ issuing orthogonally from $\Sigma$ there
exist a  constant $c_\gamma>0$  for which
\be\lb{schw}
\lim_{T\to\infty} \textrm{inf}\int^T_0 e^{-2c_\gamma\tau/(n-1)}r_\gamma(\tau)d\tau>\theta_{0\gamma}+\frac{c_\gamma}{2},\ee
where $r_\gamma(\tau)=\textrm{Ricc}(\gamma', \gamma')$ and $\theta_{0\gamma}$ the value of the expansion parameter at the intersection point between $\gamma$ and $\Sigma$. Then $\theta_\gamma$ diverges at a finite time $\tau_\gamma$ and the space $M$ is future geodesically incomplete.\\

\emph{Proof:} The proof is  a consequence of proposition 2. As discussed below (\ref{raycha}), the Raychaudhuri equation may be converted by a change of variables
$y(\tau)=-(\theta+c_\gamma)e^{-2c_\gamma\tau/(n-1)}$  into (\ref{teor}) with the functions $p(\tau)$ y $q(\tau)$ defined by (\ref{che}). The function $q(\tau)$
clearly satisfies the condition of the proposition 2, when $c_\gamma>0$. The function $p(\tau)$ will satisfy them if 
$$
\int^\tau_0 p(\tau')d\tau'>-y_0,
$$
which is translated by (\ref{che}) into 
$$
\int^\tau_0 e^{-2c_\gamma \tau'/(n-1)}\bigg(\textrm{Ricc}(\gamma', \gamma')+2\sigma^2+\frac{c_\gamma^2}{n-1}\bigg)d\tau'>\theta_{0\gamma}+c_\gamma,
$$
The last term is a simple integral whose maximal value is $c_\gamma/2$, and the second is strictly positive. Thus, it is clear that if 
$$
\lim_{T\to\infty} \textrm{inf}\int^T_0 e^{-2c_\gamma\tau/(n-1)}r_\gamma(\tau)d\tau>\theta_0+\frac{c_\gamma}{2},
$$
with $r_\gamma(\tau)=\textrm{Ricc}(\gamma', \gamma')$, then the proposition 2 applies. The last condition for quantity $r_\gamma(\tau)$, which is expressed in terms of the Ricci tensor, is exactly (\ref{schw}). Now, the proposition 2 implies that $\theta_\gamma$ explodes at a finite future time $\tau_{\gamma}$ for an space time characterized by such Ricci tensor. However, the time $\tau_{\gamma}$ depends on the chosen geodesic emanating from $\Sigma$ and, even if for every point in $\Sigma$ is finite, it does not means that there is an upper bound on it. However, when $\Sigma$ is compact, such bound exists and it can be used to show incompleteness \cite{fewster}, \cite{halo}. (Q. E. D) \\

A more explicit description of the compactness property for $\Sigma$ and can be found in \cite{halo} and we refer to the reader to that reference.
In many applications, the singularity theorems are formulated in terms of a past singularity, instead of a future one. By simple following the arguments of the previous propositions, it is elementary to find such version.\\

\emph{Generalized past singularity theorem:} Consider a globally hyperbolic space time $M$ with dimension $n>2$ and with a given compact Cauchy hyper surface 
surface $\Sigma$. Suppose that for each past directed time geodesic $\gamma: (-\infty,0]\to M$ issuing orthogonally from $\Sigma$ there
exist a  constant $c_\gamma<0$ for which
\be\lb{schw2}
\lim_{T\to-\infty} \textrm{inf}\int^0_T e^{-2c_\gamma\tau/(n-1)}r_\gamma(\tau)d\tau\geq -\theta_{0\gamma}-\frac{c_\gamma}{2},
\ee
 where $r_\gamma(\tau)=\textrm{Ricc}(\gamma', \gamma')$ and $\theta_{0\gamma}$ the value of the expansion parameter at the intersection point between $\gamma$ and $\Sigma$. Then $\theta_\gamma$ diverges at a finite past time and thus, the space $M$ is past incomplete\footnote{Note that a given space time is geodesically incomplete when there exist at least one geodesic  $\gamma: [0, a)\to M$ which can not be extended for all the values of the affine parameter $a$, that is, to $a\to \infty$. The propositions given above are in fact too restricted, since it shows that every time like geodesic is incomplete.}.
\\

In the last theorem, the condition (\ref{schw2}) implies the existence of a given fixed time $\tau_1<0$ for which
$$
\int^0_\tau e^{-2c_\gamma\tau/(n-1)}r_\gamma(\tau')d\tau' \geq -\theta_{0\gamma}-\frac{c_\gamma}{2},
$$
for any $\tau<\tau_1$. These theorems will be applied to concrete situations for $f(R)$ theories in the following. 

Note that if $r_\gamma(t)$ satisfies the strong energy condition, then the future version theorem implies that $c=0$ and $\theta_{0}(p)\leq 0$ while the past version implies that $\theta_{0}(p)\geq 0$. This is the content of the standard singularity theorems of Hawking \cite{hawking}, although this theorem is not restricted to compact hyper surfaces.

\subsection{Application to $f(R)$ theories}

The two generalized singularity theorems described in the previous are not related to any specific gravity model. In the present one, these theorems will be applied to
$f(R)$ models (\ref{eins}) with matter lagrangian ${\cal L}_m$ satisfying the strong energy conditions (\ref{sec})-(\ref{sec2}). The energy momentum tensor of the scalar field $\varphi$ is given by 
\be\lb{wi}
T_{ab}=\nabla_a \varphi\nabla_b\varphi-\frac{\widetilde{g}_{ab}}{2}\bigg(\nabla_c\varphi \nabla^c \varphi+U(\varphi)\bigg),
\ee
and such component generally does not satisfy  (\ref{sec})-(\ref{sec2}). The Einstein equations for the model (\ref{eins}) give the following expression for $r_\gamma(t)=\widetilde{R}_{ab}\gamma'^a \gamma'^b$
\be\lb{wi2}
r_\gamma(t)=8\pi \bigg((\nabla_\gamma\varphi)^2-\frac{U(\varphi)}{n-1}\bigg)+\bigg(\widetilde{T}^m_{ab}-\frac{\widetilde{g}_{ab}}{2}\widetilde{T}^m\bigg)\gamma'^a \gamma'^b.
\ee
Note that the contribution of matter to $r(t)$ is always positive due to the conditions (\ref{sec})-(\ref{sec2}). The contribution from the term $(\nabla_\gamma\varphi)^2$ is positive as well.
The only negative part may come from the term with the potential $U(\varphi)$. By taking this into account it follows that
\be\lb{ineco}
\int^T_0 e^{-\frac{2c_\gamma t}{n-1}}r(t)dt>- \frac{K^2}{2c_\gamma},
\ee
for all $T$ and $c_\gamma$. Here
$$
K=\sqrt{\frac{8\pi(n-1) U_{max}}{(n-2)}},
$$
with $U_{max}$ the maximum value that the potential takes during the evolution of $\varphi$, if this maximum exists.
Then, the conditions of the generalized singularity theorem will be satisfied when
$$
- \frac{K^2}{2c_\gamma}-\frac{c_\gamma}{2}>\theta_0.
$$
The left hand is minimized when $c_\gamma=K$, and it follows that when \be\lb{k}\theta_0<-K\ee and the Cauchy surface $\Sigma$ is compact, then the space is future geodesically incomplete.
Results of this type were considered in \cite{fewster} for GR with free scalar fields and for other type of matter as well.

The results described above apply in presence of a maximum value for the potential $U_{max}$. There are several realistic models for which this assumption is justified.
The $f(R)$ function is a free parameter of the model, but there are some physical constraints on it. The condition for the theory to be free of ghosts
is that $f'(R)>0$. The conditions for being free of tachyons is $f''(R)>0$.  In addition, there exist a large class of models in the literature for which, in the limit of large curvature $R\to \infty$
\be\lb{curdit}
f(R)\to R, \qquad f'(R)\to 1,\qquad f''(R)\to 0.
\ee
 These conditions corresponds to the scalaron $\varphi$ to have finite values as the curvature diverges. Some known examples are
\be\lb{se}
f(R)=R-R_s \beta \alpha\bigg\{1-\frac{1}{\bigg[1+\bigg(\frac{R}{R_s}\bigg)^{n}\bigg]^{\frac{1}{\beta}}}\bigg\}.
\ee
These models encodes a variety of scenarios, in particular the Starobinsky  \cite{especif3} or the Hu-Sawicki ones \cite{capoziello4}. When $n=1$ and $\beta\to\infty$ the last model reduces to \cite{maeda2}
\be\lb{jr}
f(R)=R-R_s\alpha \log\bigg(1+\frac{R}{R_s}\bigg),
\ee
which satisfies all these constraints.  Another interesting model is the following \cite{guo}
\be\lb{qcd}
f(R)=R\bigg[1-\frac{b}{1+\log\frac{R}{R_s}}\bigg].
\ee
The interest in these models is that the effective Newton $G^{eff}_N$ constant runs with the curvature in analogous
fashion as the QCD running coupling constant $g(\mu)$ with the energy scale $\mu$ of the process. 

Now, in order to apply the condition (\ref{k}) it is needed to know if $U_{max}$ exists. It is also convenient to understand if $\phi$ takes finite positive values, since this
field represents the conformal transformation between the Jordan and the Einstein frame. Let us focus in the model (\ref{se}) first.  From (\ref{dufo}) and (\ref{se}) it is directly deduced that
$$
\phi=f'(R)=1-n\alpha\frac{\bigg(\frac{R}{R_s}\bigg)^{n-1}}{\bigg[1+\bigg(\frac{R}{R_\ast}\bigg)^{n}\bigg]^{\frac{1}{\beta}+1}}.
$$
The maximum of the second term is given at the point
$$
\bigg(\frac{R}{R_s}\bigg)^{n}=\frac{\beta(n-1)}{n+1+\beta},
$$
and therefore the minimum value of $f'(R)$ is 
$$
\phi=f'_{min}(R)=1-\alpha n \frac{\bigg(\frac{\beta(n-1)}{n+1+\beta}\bigg)^{\frac{n-1}{n}}}{\bigg[1+\frac{\beta(n-1)}{n+1+\beta}\bigg]^{\frac{1}{\beta}+1}}.
$$
By choosing $\alpha$ appropriately, it may be shown that always $\phi>0$. In addition, the value of $f'(R)$ is bounded from above.
Thus $\phi$ is bounded both from above and below and is positive. This means that the conformal transformation between the Jordan and the Einstein frame is always well defined.
 On the other hand (\ref{ct}) shows that
$$
\varphi = \sqrt{\frac{3}{2k}} \log \phi,
$$
and thus $\varphi$ also takes bounded values. However, this fact does not ensure that $U(\varphi)$ is bounded, since it may have poles
for finite scalaron values.  Thus, a further check is needed in order to see that this is not the case. The expression for the potential (\ref{pot}) gives that
\be\lb{poto2}
U(\varphi) =\frac{ R f'(R)-f(R)}{2\kappa f'(R)^2},
\ee
The denominator never goes to zero, since $f'(R)>0$. The numerator is
$$
R f'(R)-f(R)=-nR_s\alpha\frac{\bigg(\frac{R}{R_s}\bigg)^{n}}{\bigg[1+\bigg(\frac{R}{R_\ast}\bigg)^{n}\bigg]^{\frac{1}{\beta}+1}}+R_s \beta \alpha\bigg\{1-\frac{1}{\bigg[1+\bigg(\frac{R}{R_s}\bigg)^{n}\bigg]^{\frac{1}{\beta}}}\bigg\}.
$$
This expression obviously has a maxima, and does not diverge for any value of the curvature when $n$ is even. Thus, $U(\varphi)$ does not have poles and reaches a maximum value.  Thus the condition (\ref{k}) make sense and, when is satisfied, it follows that the space is future geodesically incomplete.

Consider now the model (\ref{jr}). The scalar field $\phi$ is
\be\lb{jr2}
\phi=f'(R)=1-\frac{\alpha}{1+\frac{R}{R_s}},
\ee
 and is always positive when $\alpha<1$. The numerator in (\ref{poto2}) is
 $$
 R f'(R)-f(R)=-\frac{\alpha R}{1+\frac{R}{R_s}}+R_s\alpha \log\bigg(1+\frac{R}{R_s}\bigg).
 $$
 This numerator is not bounded. Thus, it can not be ensured that the generalized future singularity theorem 2 of the previous section directly applies. It is interesting to note that if the curvature $R$ is finite during the evolution, then the scalaron has positive finite values and the theorem indeed apply when $\theta_{0\gamma}<-K$. Thus, for non singular curvature values, the space may be geodesically incomplete. It is important  at this point to recall that geodesic incompleteness does not necessarily implies a curvature singularity \cite{geroch}. On the other hand, when $U_{max}\to\infty$ then the curvature $R$ is exploiting.
But it may be the case that the curvature only when $\tau\to\infty$, and this is not a point in the space time manifold. We are not in position to distinguish if this case is realized by use of our results. 
 
 Finally, for the model (\ref{qcd}) the scalar $\phi$ is
 $$
 \phi=f'(R)=1-\frac{b}{1+\log \frac{R}{R_s}}+\frac{b}{(1+\log \frac{R}{R_s})^2}.
 $$
 This scalar is not bounded from above, although it has a minima. Again, the theorem do not apply in this situation without further ad hoc assumptions.

We finish this section by noticing that the past version of this result is that when  $\theta_0>K$ the space time is past geodesically incomplete. The deduction of this fact is completely analogous to the future case.

\section{Relaxing the compactness assumption for the Cauchy surface}

The singularity results described above requires a compact Cauchy hyper surface. However, the following variation
of proposition 2 allows to circumvent this limitation \cite{fewster}.
\\

\emph{Proposition 3:} Consider an equation of the form (\ref{teor}), and suppose that  there exists a fixed time $t_1$ for which the functions $p(\tau)$ and $q(\tau)$
satisfy
\be\lb{cudo3}
\int_0^\infty \frac{d\tau}{q(\tau)}=\infty,\qquad \textrm{inf}_{T\geq 0}\int^T_0 p(\tau)d\tau+y_0=\alpha>0,
\ee
 with $q(\tau)>0$ in $[0,\infty)$.  Then, any  of the solutions $y(\tau)$ of (\ref{teor})  with the initial condition $y(0)=y_0$ exploits inside the interval $[0,\tau]$, where $\tau$ is the time defined by
 $$
 \int_0^\tau \frac{dt}{q(t)}=\frac{2}{\alpha}.
 $$
 
\emph{Proof:}  By integrating the equation (\ref{teor}) it follows that
\be\lb{teor22}
y(\tau)=\int_0^\tau \frac{y^2}{q} d\tau' +\int_0^\tau p d\tau'+y_0>\int_0^\tau \frac{y^2}{q} d\tau'+\alpha>\alpha,
\ee
the last equality follows from the positivity of $q(t)$. Now, let us introduce the quantity given by
$$
R(\tau)=\int_0^\tau \frac{y^2(\tau')}{q(\tau')} d\tau'.
$$
From (\ref{teor22}) it is clear that
\be\lb{pin}
R(\tau)=\alpha^2\int_0^\tau \frac{1}{q(\tau')} d\tau'.
\ee
As $q(\tau)$ is positive in the half positive line and $\tau>0$, it can directly be seen that this quantity is always positive.
On the other hand, an analogous reasoning to (\ref{teor22}) shows that 
\be\lb{egoc}
\frac{R^2}{q}<\dot{R}=\frac{y^2}{q},
\ee
for $\tau>0$. From here it is concluded, by fixing $\tau_2>0$, that
$$
 \int_{\tau_2}^{\tau}\frac{d\tau'}{q}<\int_{\tau_2}^{\tau}\frac{\dot{R}}{R^2}d\tau'=\frac{1}{R(\tau_2)}-\frac{1}{R(\tau)}<\frac{1}{R(\tau_2)}.
$$
By properly taking into account the inequality (\ref{pin}) it follows that
$$
\bigg(\int_{\tau_2}^{\tau}\frac{d\tau'}{q}\bigg)\bigg(\int^{\tau_2}_{0}\frac{d\tau'}{q}\bigg)\leq \frac{1}{\alpha^2},
$$
for $0\leq t_2\leq t$. Now, the intermediate value theorem allows to find a value $t_2$ for which the left hand side is
$$
\bigg(\int_{0}^{\tau}\frac{d\tau'}{q(\tau')}\bigg)^2\leq \frac{1}{\alpha^2},
$$
and this establishes the proposition. (Q. E. D)
\\

\emph{Generalized future singularity theorem 2:} Consider a globally hyperbolic space time $M$ with dimension $n>2$ and with a give Cauchy 
hyper surface $\Sigma$ for it. Suppose that for each future directed time geodesic $\gamma: [0, \infty)\to M$ issuing orthogonally from $\Sigma$ there
exist two constants $c_\gamma>0$ and $\alpha_\gamma\geq C>0$ and a time $\tau_1$ for which
\be\lb{schw22}
\textrm{inf}_{T\geq0} \int^T_0 e^{-2c_\gamma\tau/(n-1)}r_\gamma(\tau)d\tau=\alpha_\gamma+\theta_{0\gamma}+c_\gamma,\ee
for $\tau>\tau_1$, where $r_\gamma(\tau)=\textrm{Ricc}(\gamma', \gamma')$. Then $\theta_\gamma$ diverges at a finite proper time $\tau\leq (n-1)/C$ and thus, the space $M$ is future incomplete.\\

\emph{Proof:} As discussed below (\ref{raycha}), the Raychaudhuri equation may be converted by a change of variables
$y(\tau)=-(\theta+c_\gamma)e^{-2c_\gamma\tau/(n-1)}$  into (\ref{teor}) with the functions $p(\tau)$ y $q(\tau)$ defined by (\ref{che}). The function $q(\tau)$
clearly satisfies the condition of the proposition 2, when $c_\gamma>0$. The function $p(\tau)$ will satisfy them if 
$$
y_0+\textrm{inf}\int^T_0 p(\tau')d\tau=\alpha>0,
$$
which is translated by (\ref{che}) into 
$$
\textrm{inf}\int^T_0 e^{-2c_\gamma \tau'/(n-1)}\bigg(\textrm{Ricc}(\gamma', \gamma')+2\sigma^2+\frac{c_\gamma^2}{n-1}\bigg)d\tau'=\alpha_\gamma+\theta_{0\gamma}+c_\gamma,
$$
with $\alpha_\gamma>0$. Under these circumstances,  the proposition 3 applies.  Now, this proposition implies that the value of $\theta_\gamma$ will explode at a time in
the interval $[0,\tau_\gamma]$ with $\tau_\gamma$ the solution of the equation
$$
\int_{0}^{\tau_\gamma}\frac{e^{2c_\gamma t/(n-1)}}{(n-1)}dt=\frac{1}{\alpha_\gamma},
$$
 where (\ref{che}) has been taken into account. By calculating explicitly the last integral it follows that
 $$
\tau_\gamma=\frac{n-1}{2c_\gamma}\log(1+\frac{3c_\gamma}{\alpha_\gamma}).
$$
Let us rewrite the last expression as
$$
\tau_\gamma=\frac{2(n-1)}{3\alpha_\gamma}\frac{1}{x} \log(1+x).
$$
with $x=3c_\gamma/\alpha_\gamma$ a positive variable. The maximum value of the function $f(x)=x^{-1}\log(1+x)$ for positive values of $x$ is  $f(0)=1$.
Thus 
$$
\tau_\gamma\leq \frac{2(n-1)}{3\alpha_\gamma}\leq \frac{2(n-1)}{3C}.
$$
The last part of this inequality does not depend on the curve $\gamma$, thus any $\theta_\gamma$ exploits at a time less than $\tau_0=2(n-1)/C$. A direct application of proposition 1 and the previous results  shows that the space is geodesically incomplete. (Q. E. D)\\

Note that, the smaller $C$ is, the larger the  bound for the explosion time will be. In fact, the singularity theorems of the previous sections corresponds to the case $C\to 0$.
For completeness, we mention that the  past version of this theorem follows by replacing $\int_0^T$ by $\int_T^0$, $\theta_{0\gamma}$ by $-\theta_{0\gamma}$ and by considering  constant $c_\gamma<0$. The constant $\alpha_\gamma$
is, as before, positive.

\subsection{Singularities for $f(R)$ models}

In order to apply the generalized future singularity  theorem 2 given above, it should be noted in (\ref{schw2}) that when $
e^{-2c_\gamma\tau/(n-1)}r_\gamma(\tau)$ is integrable, it follows that
$$
\int^T_0 e^{-2c_\gamma\tau/(n-1)}r_\gamma(\tau) d\tau\geq \int^\infty_0 e^{-2c_\gamma\tau/(n-1)}r_{\gamma-}(\tau) d\tau,
$$
for any value of $T>0$. Here $r_{-\gamma}(\tau)=$Min$(0, r_\gamma(t))$. So,  if 
$$
\int^\infty_0 e^{-2c_\gamma\tau/(n-1)}r_{\gamma-}(\tau)d\tau=\alpha_\gamma+\theta_{0\gamma}+c_\gamma,
$$
is satisfied with $\alpha_\gamma>C>0$, the hypothesis of the generalized future theorem 2 will be satisfied.
Now, taking into account (\ref{wi}) and (\ref{wi2}) the last condition is
\be\lb{ineco2}
\int^\infty_0 e^{-\frac{2c_\gamma t}{n-1}}r_{\gamma-}(t)dt>- \frac{K^2}{2c_\gamma},
\ee
with
$$
K=\sqrt{\frac{8\pi(n-1) U_{max}}{(n-2)}},
$$
with $U_{max}$ the maximum value that the potential takes during the evolution of $\varphi$, if this maximum exists.
Then, the conditions of the generalized singularity theorem will be satisfied when
$$
- \frac{K^2}{2c_\gamma}-c_\gamma=\theta_{0\gamma}+\alpha_\gamma.
$$
The left hand is minimized when $\sqrt{2}c_\gamma=K$, thus the theorem will be satisfied when
$$
- \sqrt{2}K-\alpha_\gamma'>\theta_{0\gamma},
$$
with $\alpha_\gamma'>C>0$. The previous theorem insures that the blowup proper time is less than  $\tau_0=\frac{2(n-1)}{3C}$.

Analogous considerations hold for past singularities. The manifold will be past geodesically incomplete when
$$
\sqrt{2}K+\alpha_\gamma'<\theta_{0\gamma},
$$
where  in this case still $\alpha_\gamma'\geq C>0$. The past blowup proper time $\tau_0$ satisfy $0>\tau_0>-\frac{2(n-1)}{3C}$.

The application of this result to the previous discussed theories is straightforward. The results are completely analogous, since a singularity is formed for sure when the potential gets a maximum, as in the situation for compact Cauchy surfaces.

\section{Singularities for the Jordan metric}

\subsection{The vacuum case}

As it was shown in (\ref{eins}), there exist a conformal transformation for which the $f(R)$ models reduce to ordinary GR plus an scalar field $\varphi$, the scalaron, together with matter coupled non trivially
to the scalar degree of freedom  by a term of the form (\ref{lm}). In this context, it is of great importance to analyze whether or not the resulting system is well behaved. For instance, it may be of interest to understand under which situations, given some suitable 
boundary conditions, the scalaron $\varphi$ is uniquely defined. Otherwise, the conformal transformation to the original Jordan frame would be ambiguous. The present section deals with type of issues.

It is convenient first to study the vacuum model, that is, the case for which ${\cal L}_m=0$. In
this case, as previously stated, the theory is reduced in the Einstein frame to GR coupled to an scalar field $\varphi$. Surprisingly, it was only recently when the properties of such system were firmly established.
These properties are fully proved in \cite{ringstrom} and we refer the reader to that reference for further details. However, the main results will be exposed here without proof. 

Let ($M$, $g$) be a time oriented Lorentz manifold, together with an scalar field $\varphi$ which can be considered as an smooth function on $M$. Let $\Sigma$ an smooth space like surface
and let $h_{ij}$ and $k_{ij}$ the metric and the second fundamental form on $\Sigma$ induced by  the space time metric $g_{ij}$.  The future directed normal unit vector
to $\Sigma$ will be denoted by $N$ and  the Levi-Civita covariant derivative induced on $\Sigma$ by $h$ will be denoted by $D_\mu$. In these terms, the following equalities take place for the Einstein tensor $G_{ij}=R_{ij}-g_{ij}R$ of the space time manifold ($M$, $g$)
\be\lb{c1}
G(N_p, N_p)=\frac{1}{2}\bigg[S-k_{ij}k^{ij}+(\textrm{Tr}_g k)^2\bigg](p),
\ee
\be\lb{c2}
G(N_p, v)=\frac{1}{2}\bigg[D^j k_{ji}-D_i\textrm{Tr}_g k\bigg]v^i.
\ee
 Here $p\in \Sigma$ and $v\in T_p\Sigma$. In addition, $S$ denotes the scalar curvature of the three dimensional space like hyper surface ($\Sigma$, $h$).
 
The relations given above are quite general and the scalaron does not play any role in its deduction.  But  if it is further assumed that the system in consideration is the Einstein one coupled to an scalar field $\varphi$, it is deduced from (\ref{c1}) the so called hamiltonian constraint 
 \be\lb{c3}
\frac{1}{2}\bigg[S-k_{ij}k^{ij}+(\textrm{Tr}_g k)^2\bigg]=\rho, 
\ee
with $\rho$ defined as
$$
\rho=\frac{1}{2}\bigg[(N\varphi)^2+D_i\varphi D^i \varphi\bigg]+U(\varphi).
$$
On the other hand, the constraint (\ref{c2}) give the so called momentum constraint
\be\lb{c4}
D^j k_{ji}-D_i\textrm{Tr}_g k= j(\varphi),
\ee
with $j(\varphi)=N(\varphi)D^i \varphi$. 

Given the equations (\ref{c1})-(\ref{c4}) the initial formulation for the vacuum problem for the coupled scalar-Einstein model goes as follows. The initial data 
consist on an $n$ dimensional manifold $\Sigma$ endowed with a riemannian metric $g_0$ on it, together with a covariant tensor $k_{ij}$, and two smooth functions $\varphi_0$ and $\varphi_1$ on $\Sigma$,
 satisfying
 \be\lb{c5}
r-k_{ij}k^{ij}+(\textrm{Tr}_g k)^2=(\varphi_1)^2+D_i\varphi_0 D^i \varphi_0+2U(\varphi_0).
\ee
\be\lb{c6}
D^j k_{ji}-D_i\textrm{Tr}_g k= \varphi_1D^i \varphi_0.
\ee
Here $D^j$ is the Levi-Civita connection on $\Sigma$ and $r$ the scalar curvature of ($g_0$, $\Sigma$), and the indices are raised and lowered with the help of $g_0$.
Given this data, the problem is to find an $n+1$ manifold $M$ endowed with a Lorenzian metric $g$ and a $C^{\infty}(M)$ map $\varphi$ such that the Einstein equations are
satisfied, together with an embedding $i:\Sigma\to M$ such that $i^\ast: g\to g_0$ and $\varphi\:o \:i=\varphi_0$. If in addition $N$ is a future directed normal time like vector
to $\Sigma$ and $K$ is the second form of $i(\Sigma)$, then $i^\ast(K)=k$ and $(N\varphi)\;o\;i=\varphi_1$. The triple ($M$, $g$, $\varphi$) is known as the development of the data.
If furthermore $i(\Sigma)$ is a Cauchy hyper surface in ($M$, $g$) then ($M$, $g$, $\varphi$)  is called globally hyperbolic development.
The first question that arise is wether or not such developments exist. The answer is affirmative, as shown in the following preposition \cite{ringstrom}.
\\

\emph{Proposition 4:} There always exist a global hyperbolic development for the data ($\Sigma$, $k$, $g_0$, $\varphi_0$, $\varphi_1$) satisfying the constraints (\ref{c5})-(\ref{c6}) described above.
\\

The following proposition  shows that two different developments of a given data are an extension of a common development \cite{ringstrom}.
\\

\emph{Proposition 5:} Consider a given  data ($\Sigma$, $k$, $g_0$, $\varphi_0$, $\varphi_1$) and two hyperbolic developments ($M_a$, $g_a$, $\varphi_a$)
and ($M_b$, $g_b$, $\varphi_b$) with corresponding embeddings $i_a:  \Sigma\to M_a$ and $i_b:  \Sigma\to M_b$. Then there exist a global hyperbolic development
($M$, $g$, $\varphi$) with a corresponding embedding $i:  \Sigma\to M$ and an smooth orientation preserving maps $\psi_a: M\to M_a$ and $\psi_b: M\to M_b$, which are 
diffeomorphisms onto their images, such that $\psi_a^\ast g_a=g$, $\psi_a^\ast \varphi_a=\varphi$ and $\psi_b^\ast g_b=g$, $\psi_b^\ast \varphi_b=\varphi$. In addition 
$\psi_a\; o\: i=i_a$ and $\psi_b\; o\: i=i_b$. \\

A fundamental notion is then the notion of an maximal hyperbolic development. An  hyperbolic development ($M$, $g$, $\varphi$) is called maximal if, for any other global hyperbolic development ($M'$, $g'$, $\varphi'$),
there is an embedding $i':  \Sigma\to M'$ and an smooth orientation preserving maps $\psi: M'\to M$ 
such that $\psi^\ast g=g'$, $\psi^\ast \varphi=\varphi'$ and
$\psi\; o\: i'=i$. The following proposition shows that maximal hyperbolic developments  always exist \cite{ringstrom}.
\\

\emph{Proposition 6:} Given a data ($\Sigma$, $k$, $g_0$, $\varphi_0$, $\varphi_1$)  there exist a maximal global hyperbolic development, which is unique up to an isometry.
\\

The last proposition we would like to mention is related to the Cauchy stability of the problem \cite{ringstrom}.
\\

\emph{Proposition 7:} Let ($M=\Sigma\times I$, $g$, $\varphi$) a background solution of the Einstein-scalar system. By denoting by  ($\Sigma$, $k$, $g_0$, $\varphi_0$, $\varphi_1$) the data induced
on $\{0\}\times \Sigma$ by the full solution, consider a sequence ($k_j$, $g_{0j}$, $\varphi_{j0}$, $\varphi_{1j}$) of initial conditions converging to ($\Sigma$, $k$, $g_0$, $\varphi_0$, $\varphi_1$) in the Sobolev norm $H^{l+1}$, with 
$2l>n+2$ with $n+1$ the space time dimension, and satisfying the corresponding constraint equations. Then there exist $t_{1j}$ and $t_{2j}$ such that on $M_j=\Sigma\times (t_{1j}, t_{2j})$ 
there exist a Lorentzian metric $h_j$ and an scalar $\varphi_j$ which satisfies the combined Einstein scalar field equations, and such that the initial data is 
 ($k_j$, $g_{0j}$, $\varphi_{j0}$, $\varphi_{1j}$). The surface $\tau\times \Sigma$ is a Cauchy one when $\tau \in (t_{1j}, t_{2j})$. Furthermore, when $\tau\in I$, 
 the  data on such Cauchy hyper surface induced by $(h_j, \varphi_j)$ converges to the one induced by ($g$, $\varphi$) for large $j$.
 \\

The last statement is the most subtle of this theorem. In any case, all these results shows that the Einstein-scalar field system is a well conditioned one.
However, the analysis is more involved when matter is added. This problem is to be discussed below.

\subsection{The addition of matter}

The theorems described above apply for a vacuum solution of the theory only. The theorems are not generalized in straightforward manner when matter is added, since the addition of the lagrangian 
(\ref{lm}) changes the equations of motion for $\varphi$. This is unless the matter term is conformally invariant. In four dimensions, this is the case for the Maxwell electromagnetic lagrangian
\be\lb{max}
L_m=-\frac{1}{4\pi}F_{\mu\nu}F^{\mu\nu}.
\ee
When such type  matter is added,  there should not
be a problem coupling the scalar field to any other type of matter for which
local existence and uniqueness has been proven. Nevertheless, this has to
be checked in each individual case. 
 
For the non conformal case, the problem is more difficult, but it may be studied following the suggestions in \cite{salgado}-\cite{capo}. Following these references, one may study the theory
in the O' Hanlon formulation with the following harmonic coordinate choice
\begin{equation}\label{prosto}
F^i_{\phi}= F^i - H^i =0 \qquad {\rm with}\qquad F^i :=g^{pq}\Gamma^i_{pq}, \quad H^i = \frac{1}{\phi}\nabla^i\phi.
\end{equation} 
The equation of motions for the free case reduce to
\begin{equation}\label{novopoko}
R_{ij} = \frac{1}{\phi}\left[T_{ij} - \frac{1}{2}T g_{ij}\right],
\end{equation}
where the energy momentum tensor is
\begin{equation}\label{tei}
T_{ij}= \nabla_i\nabla_j \phi - g_{ij}g^{pq}\nabla_p\nabla_q\phi - \frac{1}{2}V(\phi)g_{ij},
\end{equation}
The Ricci tensor in this gauge takes the simplifying form
\begin{equation}\label{simply}
R_{ij} = R_{ij}^\phi + \frac{1}{2}\left[ g_{ip}\partial_j\left( F^p_\phi + H^p \right) + g_{jp}\partial_i\left( F^p_\phi + H^p \right)\right]
\end{equation}
with
\begin{equation}\label{simpy22}
R_{ij}^\phi= - \frac{1}{2}g^{pq}\partial^2_{pq} g_{ij} + A_{ij} (g,\partial g),
\end{equation}
where only  first order derivatives  appear in the functions $A_{ij}$. Assuming that $F^i_\phi =0$ and taking the expression of $H^i$ into account, we obtain the following representation
\begin{equation}\label{gleiche}
R_{ij} = - \frac{1}{2}g^{pq}\partial^2_{pq}g_{ij} + \frac{1}{\phi}\partial^2_{ij}\phi + B_{ij}(g,\phi,\partial g,\partial\phi)
\end{equation}
where the functions $B_{ij}$ depend on the metric $g$, the scalar field $\phi$ and their first order derivatives. In addition
$$
\frac{1}{\phi}\left[T_{ij} - \frac{1}{2}Tg_{ij}\right] = \frac{1}{\phi}\partial^2_{ij}\phi + C_{ij}(g,\phi,\partial g,\partial\phi)
$$
Again, in the functions $C_{ij}$, only first order derivatives are involved. The last two formulas shows that the equations of motion are
\begin{equation}\label{grava}
g^{pq}\partial^2_{pq}g_{ij} = D_{ij}(g,\phi,\partial g,\partial\phi).
\end{equation}
The initial data for these equations should satisfy the two constraints
\begin{equation}\label{greve}
F_\phi=0,\qquad G^{0i}=\frac{1}{\phi}T^{0i} \quad i=0,\ldots,3,
\end{equation}
the last one is the hamiltonian constraint. On the other hand
\begin{equation}\label{gero}
\nabla^i\left(\phi G_{ij} - T_{ij}\right) = \left(\nabla^i\phi\right)R_{ij} -\frac{1}{2}\phi_j\left( R - \frac{dV}{d\phi} \right) 
+\phi\nabla^i G_{ij}
- \left( \nabla^i\nabla_i\nabla_j 
- \nabla_j\nabla^i\nabla_i \right)\phi
\end{equation}
Now, by taking into account that $\nabla^i G_{ij}=0$, the equality
$$
(\nabla^i\phi)R_{ij} =- \left( \nabla^i\nabla_i\nabla_j 
- \nabla_j\nabla^i\nabla_i \right)\phi,
$$
and that $R=V'(\phi)$ it is obtained that
\be\lb{diverg}
\nabla^i\left(\phi G_{ij} - T_{ij}\right) =0.
\ee
This can be generalized to the case when matter fields are present, such as electromagnetic, perfect fluids and dust. In these case the authors of \cite{salgado}-\cite{capo} suggest that the Leray and Choquet-Bruhat theorems \cite{leray}-\cite{cb} can be implemented to show
that this is a well posed problem. It may be of interest to make an explicit proof of this important fact in the future.

\subsection{The singularity in the Jordan frame}

The propositions of the previous sections have shown two main results. The first is that for $f(R)$ models, in the  globally hyperbolic case with bounded values for the scalaron $\varphi$, and with suitable 
conditions on a given Cauchy surface $\Sigma$, the Einstein metric $\widetilde{g}_{\mu\nu}$ is time like geodesically incomplete. On the other hand, for such conditions, the scalaron and the metric evolution is well posed. In particular, the values of $\varphi$ and $\widetilde{g}$ at future or past times are uniquely determined by the data on the Cauchy surface. The metric in the Jordan frame is given by (\ref{ct}) and it follows that
$$
\widetilde{g}_{ab}=e^{\sqrt{\frac{2k}{3}} \varphi}g_{ab},
$$
and therefore 
$$
e^{-\sqrt{\frac{2k}{3}} \varphi_{max}}\widetilde{g}_{ab}<g_{ab}<e^{-\sqrt{\frac{2k}{3}} \varphi_{min}}\widetilde{g}_{ab}.
$$
Here $\varphi_{min}$ and $\varphi_{max}$ are the minimum and the maximum of $\varphi$ at the full evolution, which our conditions ensure to exist. The metric $\widetilde{g}$ can not be extended 
beyond certain finite time $\tau_0$ and therefore $g_{ab}$ can not be extended beyond certain time $\tau'$
$$
e^{-\sqrt{\frac{k}{6}} \varphi_{max}}\tau_0<\tau'<e^{-\sqrt{\frac{k}{6}} \varphi_{min}}\tau_0.
$$
Thus the Jordan metric is also past or future geodesically incomplete in this situation.

It is usually emphasized in the literature that the scalaron $\varphi$ as a function of the curvature $R$ is not surjective. In other words there may exist different
values of the curvature $R_1$ and $R_2$ giving rise to the same value for $\varphi$. However, this is not related to the uniqueness results described here. These results  shows instead that once the 
initial data is properly formulated on the Cauchy surface, the evolution of $\varphi$ and $\widetilde{g}$ is completely determined. Note that, in addition, the information about the curvature is given by both quantities $\widetilde{g}$ and $\varphi$, not  by $\varphi$ alone.

\section{Discussion}

In the present work, the cosmological singularity theorems of Hawking where extended to $f(R)$ models, by assuming that the matter coupled to these models satisfy the condition $\bigg(T_{ij}-\frac{g_{ij}}{2} T\bigg)k^i k^j\geq 0$ for any generic unit time like field. The difficulty relies in that this condition do not imply that $R_{ij}\gamma^i\gamma^j\geq 0$ for every unit time like vector field,  in the context of $f(R)$ models. The last is a key property in proving the standard singularity theorems. Furthermore, the additional degree of freedom, the scalaron, does not respect the condition $\bigg(T_{ij}-\frac{g_{ij}}{2} T\bigg)k^i k^j\geq 0$ either. These complications were sorted out by use of certain generic results about non Lipschitz differential equations.

The application of theorems presented here is straightforward  for certain $f(R)$ models such as the Hu-Sawicki or the Starobinsky ones \cite{especif3}, \cite{capoziello4}. For these scenarios, the corresponding scalaron potential has a maximum, which is one of the requirements for the singularity to occur. For other models, this condition is not automatically satisfied and should be checked individually, by considering different types of matter couplings.  The negation of the hypothesis of these propositions presented here does not necessarily  imply that a singularity is absent. In other word, we do not claim that we have exhausted all the possible situations for which a singularity take place, but just a class of them.

An interesting task may be to relax the global hyperbolicity condition for the underlying space time.  In addition, it may be of interest as well to generalize the Penrose singularity theorems \cite{penrose} to these scenarios. We leave these matters for a future investigation. 
\\

{\bf Acknowledgements:}  A discussion with Guillermo Silva, Nicol\'as Grandi, Mauricio Sturla y W. Helmut Baron are acknowledged. O. S is supported by CONICET (Argentina).


\begin{thebibliography}{99}

\bibitem{acelerado} S. Perlmutter et al. Nature 391, 51 (1998); A. Riess et al.,
 Astron. J. 116 1009,1998.

\bibitem{quintessence}  B. Ratra and P.J.E. Peebles, Phys. Rev. D 37, 3406 (1988); P.J. E.
Peebles and B. Ratra, Astrophys. J. 325, L17 (1988); K. Coble, S. Dodelson, and
J.A. Frieman, Phys. Rev. D 55, 1851 (1997)

\bibitem{quintessence3} M.S. Turner and M.
White, Phys. Rev. D 56, 4439 (1997); R.R. Caldwell, R. Dave and
P.J. Steinhardt, Phys. Rev. Lett. 80, 1582 (1998).

\bibitem{quintessence2} C.T. Hill, D.N. Schramm, and J.N. Fry, Comments Nucl. Part. Phys. 19, 25 (1989); J.
Frieman, C. Hill, and R. Watkins, Phys. Rev. D 46, 1226 (1992); J. Frieman, C. Hill,
A. Stebbins, and I. Waga, Phys. Rev. Lett. 75, 2077 (1995).

\bibitem{cosmological constant} S.M. Carroll, W.H. Press, and E.L. Turner, Ann. Rev. Astron. Astrophys. 30, 499
(1992).

\bibitem{rotation} J. Binney and S. Tremaine, Galactic dynamics, Princeton, Princeton University Press, (1987); M. Persic, P. Salucci and F. Stel, Month. Not. R. Astron. Soc. 281, 27 (1996); A. Borriello and P. Salucci, Month. Not. R. Astron. Soc. 323, 285 (2001).

\bibitem{wimp} . M. Overduin and P. S. Wesson, Phys. Rept. 402, 267.


\bibitem{wimp2} C. G. Bohmer and
T. Harko, Month. Not. R. Astron. Soc. 379, 393 (2007): S. J. Sin, Phys. Rev. D50, 3650 (1994); M. P. Silverman and R. L. Mallett, Gen. Rel. Grav. 34, 633 (2002)

\bibitem{wimpo} M. Nishiyama, M. Morita and M. Morikawa, astro-ph-0403571 (2004); F. Ferrer and J. A. Grifols, JCAP 0412, 012 (2004); C. G. Bohmer and T. Harko, JCAP 06, 025 (2007).

\bibitem{wimp3} I. F. M. Albuquerque and L. Baudis, Phys. Rev. Lett. 90, 221301 (2003).

\bibitem{will} C. Will Living Rev. Rel. 9, 3 ,2005.
\bibitem{especif2} S. M. Carroll, V. Duvvuri, M. Trodden and M. S. Turner, Phys. Rev. D 70, 043528 (2004).
\bibitem{especif3} A. A. Starobinsky, Phys. Lett. B 91, 99 (1980).

\bibitem{cosmos1} L. Amendola, D. Polarski and S. Tsujikawa, Phys. Rev.
Lett. 98, c131302 (2007).
\bibitem{cosmos2}S. Capozziello, S. Nojiri, S. D. Odintsov and A. Troisi,
Phys. Lett. B639, 135 (2006).
\bibitem{cosmos3} S. Nojiri and S. D. Odintsov, Phys. Rev. D74, 086005
(2006); S. Nojiri, S. D. Odintsov and P. V. Tretyakov, arXiv:0704.2520 (2007); S. Nojiri and S. D. Odintsov, arXiv:0706.1378 (2007).


\bibitem{capoziello3}  M. Amarzguioui, O. Elgaroy, D. F. Mota and T. Multa- maki, Astron. Astrophys. 454, 707 (2006); L. Amendola, R. Gannouji, D. Polarski and S. Tsujikawa, Phys. Rev. D75, 083504 (2007)

\bibitem{capu} A. A. Starobinsky, 0706.2041 [astro-ph] (2007); B. Li, J. D. Barrow and D. F. Mota, 0705.3795 [gr-qc]; S. Tsujikawa, arXiv: 0709.1391 [astro-ph] (2007).
\bibitem{capoziello4}  W. Hu and I. Sawicki, Phys. Rev. D76, 064004 (2007), 
\bibitem{capoziello5}  S. E. Perez Bergliaffa, Phys. Lett. B642, 311 (2006); J. Santos, J. S. Alcaniz, M. J. Reboucas and F. C. Carvalho, arXiv:0708.0411 [astro-ph] (2007).
\bibitem{capuziello6} G. Cognola, E. Elizalde, S. Nojiri, S. D. Odintsov and
S. Zerbini, JCAP 0502, 010 (2005); V. Faraoni, Phys. Rev. D72, 061501 (2005).

\bibitem{capozziello6} V. Faraoni, Phys. Rev. D72, 124005 (2005); S. Nojiri and S. D. Odintsov, Int. J. Geom. Meth. Mod. Phys. 4, 115 (2007);
V. Faraoni, Phys. Rev. D75, 067302 (2007).

\bibitem{dick1} R. Dick, Gen. Rel. Grav. 36, 217 (2004), gr-qc-0307052.
\bibitem{dick2} A. D. Dolgov and M. Kawasaki, Phys. Lett. B573, 1 (2003),
astro-ph-0307285.

\bibitem{dick3} T. Clifton and J. D. Barrow, Phys. Rev. D72, 103005 (2005),
gr-qc-0509059.
\bibitem{dick4} S. M. Carroll and M. Kaplinghat, Phys. Rev. D65, 063507
(2002), astro-ph-0108002.


\bibitem{especif1} H. A. Buchdahl, Mon. Not. Roy. Astron. Soc. 150, 1 (1970); J. D. Barrow and A. C. Ottewill, J. Phys. A: Math. Gen. 16, 2757 (1983).

\bibitem{capo2} C. G. Bohmer, L. Hollenstein and F. S. N. Lobo, to appear in Phys. Rev. D, 0706.1663 [gr- qc] (2007).

\bibitem{cama1} J. Khoury and A. Weltman, Phys. Rev. Lett. 93, 171104 (2004),
astro-ph-0309300.
\bibitem{cama2} J. Khoury and A. Weltman, Phys. Rev. D69, 044026 (2004),
astro-ph-0309411.

\bibitem{cama3} P. Brax, C. van de Bruck, A.-C. Davis, J. Khoury, and A. Weltman, Phys. Rev. D70, 123518 (2004), astro-ph-0408415.

\bibitem{tegmark} Thomas Faulkner, Max Tegmark, Emory F. Bunn, Yi Ma Phys. Rev. D 76 063505,2007.

\bibitem{capoziello7} T. Chiba, Phys. Lett. B575, 1 (2003); A. L. Erick- cek, T. L. Smith and M. Kamionkowski, Phys. Rev. D74, 121501 (2006); S. Nojiri and S. D. Odintsov, arXiv:0708.0924 [hep-th] (2007).

\bibitem{capoziello8}T. Chiba, T. L. Smith and A. L. Erickcek, Phys. Rev. D75, 124014 (2007).
\bibitem{capoziello9} V. Faraoni and N. Lanahan-Tremblay, arXiv:0712.3252.
\bibitem{capoziello10} G. J. Olmo, Phys. Rev. D75, 023511 (2007).
\bibitem{capoziello11} S. Nojiri and S. D. Odintsov, Phys. Rev. D68, 123512 (2003); V. Faraoni, Phys. Rev. D74, 023529 (2006); T. Faulkner, M. Tegmark, E. F. Bunn and Y. Mao, Phys.
Rev. D76, 063505 (2007)
\bibitem{capoziello12}I. Sawicki and W. Hu, Phys. Rev. D75, 127502 (2007).
\bibitem{ferraro} R. Ferraro AIP Conf. Proc. 1471, 103 (2012).
 \bibitem{capoziello13} L. Amendola and S. Tsujikawa, 0705.0396 [astro-ph]
(2007).

\bibitem{cut1} Y. Sobouti, Astron. Astrophys. 464, 921 (2007).
\bibitem{cut2}  R. Saffari and S. Rahvar, arXiv:0708.1482 (2007).
\bibitem{cut3}  O. Bertolami, C. G. B?ohmer, T. Harko and F. S. N. Lobo,
Phys. Rev. D75, 104016 (2007); O. Bertolami and
J. Paramos, arXiv:0709.3988 [astro-ph] (2007).
\bibitem{cut4} S. Capozziello, V. F. Cardone and A. Troisi, JCAP 0608, 001 (2006); S. Capozziello, V. F. Cardone and A. Troisi,
Mon. Not. R. Astron. Soc. 375, 1423 (2007).
\bibitem{cut5}  A. Borowiec, W. Godlowski and M. Szydlowski, Int. J.
Geom. Meth. Mod. Phys. 4 183 (2007).
\bibitem{cut6}  C. F. Martins and P. Salucci, arXiv:astro-ph-0703243
(2007).
\bibitem{cut7}  C. G. Bohmer, T. Harko and F. S. N. Lobo,
arXiv:0709.0046 [gr-qc] (2007).

\bibitem{cut8} Christian G. Boehmer, Tiberiu Harko, Francisco S. N. Lobo JCAP 0803 024, 2008.

\bibitem{frolov1} Frolov Phys. Rev. Lett. 101 061103, 2008. 

\bibitem{frolov3} S. Appleby and R. Battye JCAP 0805 019, 2008.

\bibitem{frolov5} S. Appleby, R. Battye, A. Moss Phys. Rev. D 81 081301, 2010


\bibitem{frolov4} S. Appleby, R. Battye, A. Starobinsky JCAP 1006:005,2010

\bibitem{frolov2}  Abha Dev, D. Jain, S. Jhingan, S. Nojiri, M. Sami, I. Thongkool Phys. Rev. D 78 083515, 2008.

\bibitem{maeda} T. Kobayashi and K. Maeda Phys. Rev. D 78 064019, 2008; Phys.Rev. D 79 024009, 2009. 

\bibitem{maeda2} V. Miranda, S. Jor‡s, I. Waga and M. Quartin Ð Phys. Rev. Lett  10,  221101, 2009.

\bibitem{guo} J. Guo and A. Frolov Phys. Rev. D 88, 124036 (2013).

\bibitem{geroch} R. Geroch Annals of Physics 48, 3 526 (1968).

\bibitem{hawking} S. Hawking and G. Ellis "The Large Scale Structure
of Space-time" Cambridge: Cambridge University Press (1973); S. Hawking and R. Penrose Proceedings of the Royal Society London
A314, 529(1970).

\bibitem{penrose} R. Penrose Phys. Rev. Lett. 14, 57 (1965).

\bibitem{fewster} C. Fewster and G. Galloway Class. Quantum Grav. 28 (2011) 125009.

\bibitem{ringstrom} H. Ringstrom "The Cauchy problem in General Relativity" European Mathematical Society (2009).

\bibitem{salgado} M. Salgado, 2006, Class. Quantum Grav., Vol. 23, 4719.

\bibitem{capo} S. Capozziello and S. Vignolo "The Cauchy problem for f(R)-gravity: an overview" arXiv:1103.2302.

\bibitem{leray} J. Leray "Hyperbolic differential equations" Institute for Advanced Study Pub., Princeton (1953).

\bibitem{cb} Y. ChoquetÐBruhat "General Relativity and the Einstein equations", Oxford University Press Inc., New York (2009).

\bibitem{wald} R. Wald General Relativity Univ. of Chicago Press (1984). 

\bibitem{dadum3} S. Capozziello, M. De Laurentis, S. Nojiri
and S.D. Odintsov Phys. Rev. D 79 (2009) 124007.


 \bibitem{dadum1} K. Bamba, S. Nojiri, S. Odintsov JCAP 0810 (2008) 045.
 
 
\bibitem{dadum2} S. Nojiri, S. Odintsov Phys.Rept. 505 (2011) 59.

\bibitem{halo} G. Galloway Math. Proc. Camb. Phil. Soc (1986) 99367.
\end{thebibliography}
\end{document}